\documentclass[prc,aps,twocolumn,showpacs]{revtex4}
\usepackage{graphicx}
\usepackage{amsbsy}
\def\pucla{$^1$}
\def\pgent{$^2$}
\def\pgatch{$^3$}
\def\pgwu{$^4$}
\def\pabln{$^5$}
\def\pregn{$^6$}
\def\pmary{$^7$}
\def\pkarl{$^8$}
\def\pzagreb{$^9$}
\def\pone{$^1$}

\begin{document}

\title{
 Measurement of $K^- p$ radiative capture to $\gamma \Lambda$
 and $\gamma \Sigma^0$ for $p_{K^-}$ between 514 and 750~MeV/$c$
}

\author{
 S.~Prakhov,\pucla\footnote[1]{Electronic address: prakhov@ucla.edu} 
 ~P.~Vancraeyveld,\pgent\footnote[2]{Electronic address: Pieter.Vancraeyveld@UGent.be}
 ~N.~Phaisangittisakul,\pucla 
 ~B.~M.~K.~Nefkens,\pucla
 ~V.~Bekrenev,\pgatch 
 ~W.~J.~Briscoe,\pgwu
 ~L.~De~Cruz,\pgent
 ~D.~Isenhower,\pabln
 ~N.~Knecht,\pregn\footnote[3]{Present address: Physics Dept. of
   University of Toronto, 60 St.George St., Toronto, Ontario,
   Canada, M5S 1A7} 
 ~A.~Koulbardis,\pgatch
 ~N.~Kozlenko,\pgatch
 ~S.~Kruglov,\pgatch 
 ~G.~Lolos,\pregn
 ~I.~Lopatin,\pgatch
 ~A.~Maru\v{s}i\'{c},\pucla\footnote[4]{Present address: Collider-Accelerator 
   Dept., Brookhaven National Laboratory, Upton, NY 11973, USA.}
 ~S.~McDonald,\pucla\footnote[5]{Present address: TRIUMF, 4004 Wesbrook Mall,
   Vancouver, B.C., Canada, V6T 2A3.}
 ~Z.~Papandreou,\pregn
 ~D.~Peaslee,\pmary\footnote[6]{Deceased}
 ~J.~W.~Price,\pucla
 ~J.~Ryckebusch,\pgent
 ~M.~Sadler,\pabln
 ~A.~Shafi,\pgwu 
 ~A.~Starostin,\pucla
 ~H.~M.~Staudenmaier,\pkarl
 ~I.~I.~Strakovsky,\pgwu
 ~I.~Supek,\pzagreb
  ~and
 ~T.~Van~Cauteren\pgent
}

\affiliation{
\pucla University of California Los Angeles, Los Angeles,
 California 90095-1547, USA}
\affiliation{
\pgent Ghent University, 9000 Ghent, Belgium}
\affiliation{
\pgatch Petersburg Nuclear Physics Institute, Gatchina 188350, Russia}
\affiliation{
\pgwu The George Washington University, Washington, D.C. 20052-0001, USA}
\affiliation{
\pabln Abilene Christian University, Abilene, Texas 79699-7963, USA}
\affiliation{
\pregn University of Regina, Saskatchewan, Canada, S4S OA2}
\affiliation{
\pmary University of Maryland, College Park, Maryland 20742-4111, USA}
\affiliation{
\pkarl Universit\"at Karlsruhe, Karlsruhe 76128, Germany}
\affiliation{
\pzagreb Rudjer Boskovic Institute, 10000 Zagreb, Croatia}

\date{\today}
         
\begin{abstract}
 Differential cross sections
 for $K^-$ radiative capture in flight on the proton,
 leading to the $\gamma\Lambda$ and $\gamma\Sigma^0$
 final states, have been measured
 at eight $K^-$ momenta between 514 and 750~MeV/$c$.
 The data were obtained with
 the Crystal Ball multiphoton spectrometer installed
 at the separated $K/\pi$ beam line C6
 of the BNL Alternating Gradient Synchrotron.
 The results  substantially improve
 the existing experimental data available
 for studying radiative decays of excited hyperon states.
 An exploratory theoretical analysis is performed within
 the Regge-plus-resonance approach.
 According to this analysis, the $\gamma\Sigma^0$ final state
 is dominated by hyperon-resonance exchange and hints
 at an important role for a resonance
 in the mass region of 1700~MeV.
 In the $\gamma\Lambda$ final state, on the other hand,
 the resonant contributions account for only half the strength,
 and the data suggest the importance of a resonance
 in the mass region of 1550~MeV.
 
\end{abstract}

\pacs{25.80.Nv, 13.75.Jz, 13.30.Ce, 14.20.Jn, 11.55.Jy}

\maketitle

\section{Introduction}

 The experimental study of the radiative reactions
 $K^-p\to \gamma\Lambda$ and $K^-p\to \gamma\Sigma^0$ is
 of special interest because these reactions are among
 the very few cases in which information about the
 radiative decays of hyperon resonances can be obtained.
 Measuring the properties
 of these resonances constitutes important input to models
 that attempt to describe the internal structure of hadrons.
 The Particle Data Group's Review of
 Particle Physics (RPP)~\cite{PDG}
 lists a number of established $\Lambda$ and $\Sigma$ resonances
 (see Table~\ref{tab:hyperonResonances}),
 albeit with large uncertainties in the masses, widths,
 and branching ratios. So far, sufficient experimental data
 and many models exist for the radiative capture of kaons
 at rest (see Refs.~\cite{Whitehouse89,Siegel95}
 and references therein), which is dominated by
 the $\Lambda(1405)S_{01}$ resonance.
 A more comprehensive study of the hyperon spectrum is possible
 through the study of in-flight capture of kaons by a proton,
 for which only a meager data set is available.
 To date, no dedicated model calculations exist
 for this reaction channel. Through crossing
 symmetry~\cite{Ji88,Williams90,Williams91,David96,VanCauteren09},
 the radiative capture process is
 intimately related to kaon photoproduction,
 which is better studied experimentally and theoretically.

 In contrast to kaon photoproduction, the measurement of
 the radiative capture in flight requires a photon
 spectrometer and careful subtraction of the large
 background from single $\pi^0$ production.
 An experimental study of the reactions
 $K^- p\to \mathrm{neutrals}$ became feasible with
 the Crystal Ball (CB) multiphoton spectrometer installed
 at the separated $K/\pi$ beam line C6
 of the BNL Alternating Gradient Synchrotron (AGS). 
 Besides the radiative processes, this study
 also involved the measurement of the following final states:
 $\eta \Lambda$, $\bar{K}^0 n$, $\pi^0\Lambda$, $\pi^0\Sigma^0$,
 $\pi^0\pi^0\Lambda$, $\pi^0\pi^0\Sigma^0$,
 and $\pi^0\pi^0\pi^0\Lambda$, reported in
 Refs.~\cite{k0sn_lpi0_spi0,etalam,l2pi0,s2pi0,l3pi0}.
 A reliable measurement of the radiative reactions
 would not be possible without 
 the extensive measurement of most of the above reactions.

 In the present work, we report on the first measurement of
 the differential cross sections for
 the radiative reactions $K^- p \to \gamma\Lambda$
 and $K^- p \to \gamma\Sigma^0$
 at eight incident $K^-$ momenta between 514 and 750~MeV/$c$.
 An exploratory theoretical analysis is performed within
 the Regge-plus-resonance
 approach~\cite{Corthals06,Corthals07_1,Corthals07_2,Vancraeyveld09}.
 Many experimental details have already been
 presented in Ref.~\cite{k0sn_lpi0_spi0}
 and are omitted in this paper.
 The preliminary results for the radiative reaction
 $K^- p \to \gamma\Lambda$ with many experimental details
 are given in Ref.~\cite{NakPhD}.
 An independent analysis of the $K^- p \to \gamma\Sigma^0$
 reaction, using the same data set,
 was recently presented in Ref.~\cite{SGam},
 reporting total cross sections or their upper limits only.
 
\begin{table*}
\caption
[tab:hyperonResonances]{
 Selection of established $\Lambda$ and $\Sigma$ resonances
 relevant to the data presented in the present work.
 We list the resonances' mass and total decay width ranges
 as given in the RPP~\protect\cite{PDG},
 in addition to their star status. In the last two columns,
 we tabulate predictions by the
 Bonn constituent-quark model~\cite{VanCauteren05,VanCauteren07}
 for the partial electromagnetic decay widths
 to the ground-state $\Lambda(1116)$ and $\Sigma^0(1193)$.
 } \label{tab:hyperonResonances}
\begin{ruledtabular}
\begin{tabular}{|c|c|c|c|c|c|c|}
\hline
  Resonance & $L_{I\cdot2J}$ & status & Mass (MeV) & Width (MeV)
 & $\Gamma_{\Lambda\gamma}$ (MeV) & $\Gamma_{\Sigma^0\gamma}$ (MeV) \\
\hline
  $\Lambda(1520)$ &$D_{03}$ & ${\ast\ast\ast\ast}$ & $1519.5\pm1.0$
 &$15.6\pm1.0$ & $0.258$ & $0.157$ \\
  $\Lambda(1600)$ &$P_{01}$ & ${\ast\ast\ast}$ & $1560-1700$ & $50-250$
 & $0.104$ & $0.0679$ \\
  $\Sigma(1660)$ & $P_{11}$ & ${\ast\ast\ast}$ & $1630-1690$ & $50-70$
 & $0.451$ & $0.578$ \\
  $\Lambda(1670)$ & $S_{01}$ & ${\ast\ast\ast\ast}$ & $1660-1680$ & $25-50$
 & $0.159\cdot10^{-3}$ & $3.827$ \\
  $\Sigma(1670)$ & $D_{13}$ & ${\ast\ast\ast\ast}$ & $1665-1685$ & $40-80$
 & $1.457$  &$0.214$ \\
  $\Lambda(1690)$ & $D_{03}$ & ${\ast\ast\ast\ast}$ & $1685-1695$ & $50-70$
 & $0.0815$ & $1.049$ \\
\hline
\end{tabular} 
\end{ruledtabular}
\end{table*} 

\section{Experimental setup}

 The experimental study was performed with
 the Crystal Ball multiphoton spectrometer, which is
 a highly segmented sphere made of NaI(Tl).
 The CB consists of 672 almost identical crystals packed in two
 hermetically sealed and evacuated hemispheres.
 The solid angle covered by the CB is 93\% of $4\pi$ steradian.
 The crystals have the shape of a truncated triangular pyramid,
 all pointed toward the center of the CB.
 The crystal length is 40.6~cm,
 which corresponds to 15.7 radiation lengths.
 The typical energy resolution for electromagnetic showers in the CB
 was $\Delta E/E = 0.020/(E[\mathrm{GeV}])^{0.36}$.
 The directions of the photon showers were measured with a resolution
 in $\theta$ (the polar angle with respect to the beam axis)
 of $\sigma_\theta = 2^\circ - 3^\circ$.
 The resolution in
 azimuthal angle $\phi$ is $\sigma_\theta/\sin\theta$. 

 The C6 line of the AGS provided a beam of negative kaons
 and pions with the $K^-/\pi^-$ ratio enhanced to about 0.1
 by two electrostatic separators.
 The beam particles were incident on a 10-cm-long
 liquid-hydrogen (LH$_2$) target located in the center
 of the Crystal Ball.
 The momentum resolution $\sigma_p/p$ for an individual
 incident kaon varied from 0.6\% to 1.\%, depending on the momentum value.
 The mean value $p_{K^-}$ for
 the incident-momentum spectra 
 and the momentum spread $\delta_p$, which were determined
 at the target center, are listed in Table~\ref{tab:events}.
 The uncertainty in determining the mean beam momentum is $2-3$~MeV/$c$.

 The LH$_2$ target was surrounded by a 16-cm-diameter pipe made
 of four thin scintillation counters that functioned as a veto
 for the beam interactions with charged particles in the final state.
 The 120-cm length of the veto counters ensured almost
 100\% rejection of those events. The 5-mm thickness
 of the counters implied both a good efficiency in vetoing
 charged particles and a low probability
 for photon conversion. 

\section{Data handling}

 Since the Crystal Ball detector is designed as a
 multiphoton spectrometer, $K^- p$ interactions
 were studied by measuring the photons and, when possible,
 the neutron in the final state.
 The $\Lambda$ and $\Sigma^0$ hyperons were
 measured in the CB via the decay chains
 $\Lambda \to \pi^0 n \to 2\gamma n$ and
 $\Sigma^0 \to \gamma\Lambda \to \gamma\pi^0 n \to 3\gamma n$.
 As the final-state photons produce electromagnetic showers in the NaI(Tl)
 crystals, they can be recognized as so-called clusters in the density
 of the energy deposited in the CB.
 The outgoing neutrons can also be detected if the products
 of their interactions in the NaI(Tl) material produce
 enough ionization to form a cluster. 
 In general, a cluster in the CB is defined as
 a group of neighboring crystals in which energy is 
 deposited from the interaction of a photon,
 a charged particle, or a neutron produced in the
 final state.
 The software threshold for the cluster energy was chosen to be 20~MeV;
 this value optimizes the yield of the
 reconstructed events for the $K^- p\to \mathrm{neutrals}$ processes.
 
 The kinematic-fitting technique was used to select
 candidates for the radiative reactions that were studied.
 Those candidates were selected
 by testing the following hypotheses:
\begin{equation}
   K^- p \to \gamma \Lambda \to \gamma\pi^0 n \to 3\gamma n~,
\label{eqn:lgam}
\end{equation}
\begin{equation}
 K^-p\to \gamma\Sigma^0\to \gamma\gamma\Lambda\to \gamma\gamma\pi^0 n \to 4\gamma n~.
 \label{eqn:sgam}
\end{equation}
 The incident kaon was parametrized in the kinematic fit
 by the five measured variables: momentum,
 angles $\theta_x$ and $\theta_y$, and position coordinates
 $x$ and $y$ at the target.
 A photon cluster was parametrized by the three measured variables:
 energy and angles $\theta$ and $\phi$. 
 As the data were taken with a 10-cm-long LH$_2$
 target, the $z$ coordinate of the vertex was a free variable
 of the kinematic fit. Including $z$ into the fit
 improves the angular resolution in the photon directions.
 When the final-state neutron was not detected, its
 energy and two angles were free variables of
 the fit. For the neutron detected in the CB,
 the angles of its cluster
 were used as the measured variables,
 while the neutron energy was a free variable of the fit.
 Since all reactions in our analysis had a particle decaying
 in flight, the decay length of this particle was also
 a free variable of the kinematic fit.
 The corresponding secondary vertex is then
 determined by the primary-vertex coordinates, the direction
 of the decaying particle, and the decay length.

 The candidates for reaction~(\ref{eqn:lgam})
 were searched for in three- and four-cluster events.
 The three-cluster events were tested for
 the case when only the three photons were detected in the CB.
 The four-cluster events were tested for the case in which all four
 final-state particles were detected. Similarly, the candidates for
 reaction~(\ref{eqn:sgam}), which have four photons and a neutron
 in the final state, were searched for in four- and five-cluster
 events. The test of each hypothesis involved all possible permutations
 of assigning the detected clusters to the particles in the reaction chain.   
 The events for which at least one permutation satisfied
 the tested hypothesis at the 5\% confidence level, CL
 (i.e., with a probability larger than 5\%)
 were accepted as the reaction candidates.
 The permutation with the largest CL was used
 to reconstruct the kinematics of the reaction.

 The candidates selected for the two radiative reactions
 are contaminated with background events
 that have to be subtracted from the experimental distributions.
 The first source of the background events appears from processes
 that are not kaon interactions in the LH$_2$ target.
 The major fraction of these interactions are
 $K^-$ decays in the beam. This background was investigated
 using the data taken when the target was empty.
 The fraction of this background in our event candidates was
 determined from the ratio of the total number
 of the beam kaons incident on the full target
 to the corresponding number for the empty target.
 The fraction of the so-called empty-target
 background, remaining in the radiative-reaction candidates
 after all selection cuts, is at the level of $15\%-22\%$.
 The properly weighted empty-target spectra
 were subtracted from the full-target distributions before
 the acceptance correction of the latter.
 The second source of the background events appears from
 the following processes,
\begin{equation}
   K^- p \to K^0_S n \to \pi^0 \pi^0 n \to 4\gamma n~,
\label{eqn:k0sn}
\end {equation} 
\begin{equation}
   K^- p \to \pi^0 \Lambda \to \pi^0 \pi^0 n \to 4\gamma n~,
\label{eqn:lpi0}
\end{equation}
\begin{equation}
 K^-p\to \pi^0\Sigma^0\to \pi^0\gamma\Lambda\to \pi^0\gamma\pi^0 n \to 5\gamma n~,
 \label{eqn:spi0}
\end{equation}
 which can be misidentified as the radiative reactions
 when some of the final-state photons are not detected in the CB.
 Our measurement of these three reactions is given in detail
 in Ref.~\cite{k0sn_lpi0_spi0}. Based on the results
 from Ref.~\cite{k0sn_lpi0_spi0}, the background from each of the
 three reactions can be understood via a Monte Carlo simulation.
 
 A Monte Carlo (MC) simulation of reactions~(\ref{eqn:lgam})
 and (\ref{eqn:sgam}) was used to determine
 their acceptance. Based on the MC simulation of
 reactions~(\ref{eqn:k0sn}), (\ref{eqn:lpi0}),
 and (\ref{eqn:spi0}), their remaining backgrounds
 were subtracted from the radiative-reaction candidates.
 Reactions~(\ref{eqn:lgam}) and (\ref{eqn:sgam}) were simulated
 with an isotropic production-angle distribution.
 Reactions~(\ref{eqn:k0sn}), (\ref{eqn:lpi0}), and (\ref{eqn:spi0})
 were simulated according to their differential cross sections
 measured using the same data~\cite{k0sn_lpi0_spi0}.
 The MC events of each beam momentum were then 
 propagated through a full {\sc GEANT} (version 3.21)
 simulation of the CB detector, folded with the CB resolutions and
 trigger conditions, and analyzed the same way as the experimental data.
 The small difference between the experimental data
 and the MC simulation for the neutron response
 in the CB was not important, 
 as we summed the events with and without the neutron detected.

 The analysis of the MC simulation for
 the background reactions showed that $\pi^0\Lambda$ events
 cause the largest contamination
 of the $\gamma\Lambda$ candidates.
 This occurs when one photon from the decay of the outgoing $\pi^0$
 is not detected in the CB. Similarly, $\pi^0\Sigma^0$ events
 constitute the largest contamination
 of the $\gamma\Sigma^0$ candidates.
 In four- and five-cluster events,
 the background from reactions~(\ref{eqn:k0sn}),
 (\ref{eqn:lpi0}), and (\ref{eqn:spi0}) was partially suppressed
 by the 2\% CL cut on the hypotheses of these reactions
 themselves; i.e., all events that satisfied these hypotheses
 with probability larger than 2\% were rejected.
 Further suppression of the remaining background events can
 be done by tightening the cut on the CL of the hypotheses
 used to select the radiative-reaction candidates.
 However, it leads to decreasing statistics.
 The MC spectra with the remaining background
 from reactions~(\ref{eqn:k0sn}),
 (\ref{eqn:lpi0}), and (\ref{eqn:spi0}) were then subtracted from
 the experimental distributions with the radiative candidates.
 The normalization of the spectra for the subtraction was based
 on the ratio between the number of the simulated events for
 the background reactions to the number of the experimental
 events found for those reactions themselves in the data
 (see Ref.~\cite{k0sn_lpi0_spi0} for details).    

\begin{figure}
\includegraphics[width=8.0cm,height=8.5cm,bbllx=0.5cm,bblly=1.0cm,bburx=19.5cm,bbury=19.5cm]{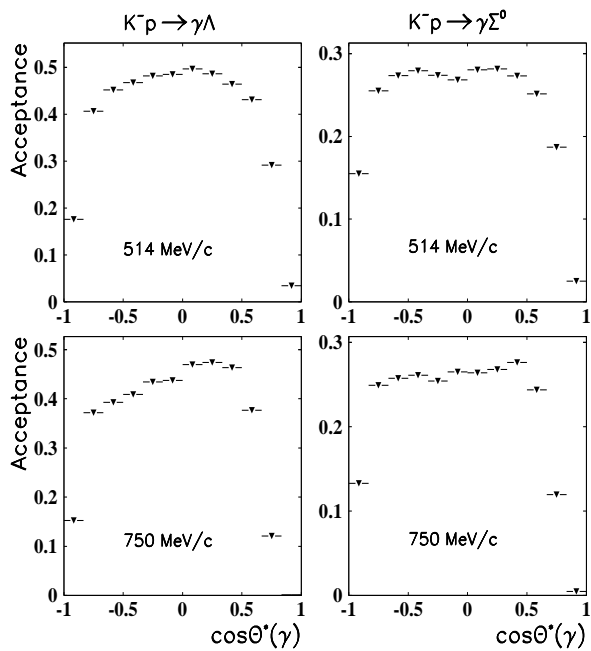}
\caption{
 Acceptance for the production angle $\theta^*$ of the outgoing photon
 in the c.m. frame; it is shown for the reactions
 $K^-p \to \gamma \Lambda$ and $K^-p \to \gamma \Sigma^0$ at beam
 momenta of 514 and 750~MeV/$c$.  
}
 \label{fig:lgam_sgam_dxs_acc} 
\end{figure}

 The acceptance for both reactions
 $K^-p \to \gamma \Lambda$ and $K^-p \to \gamma \Sigma^0$ was
 determined as a function of $\cos\theta^*$,
 where $\theta^*$ is the production angle of
 the outgoing photon (i.e., the angle between the direction
 of the outgoing photon and the incident $K^-$ meson) 
 in the center-of-mass (c.m.) frame.
 The acceptance for both the radiative reactions is 
 shown in Fig.~\ref{fig:lgam_sgam_dxs_acc} for two beam momenta:
 514 and 750~MeV/$c$. These acceptances include the effects
 of all our standard cuts used in the event selection.
 A poorer $\cos\theta^*$ acceptance of the forward $\theta^*$ angles
 for a higher beam momentum
 is mostly due to the larger threshold on the CB total energy
 in the event trigger.
 This threshold was 0.9~GeV for $p_{K^-} = 514$~MeV/$c$
 and became 1.5~GeV for $p_{K^-} = 750$~MeV/$c$.
 
\section{Experimental results and their interpretation}

 The number of the radiative-reaction candidates initially selected
 with our standard cuts 
 and the number of the experimental events left after
 the background subtraction are listed in Table~\ref{tab:events}
 for our two radiative reactions at the eight beam momenta.
 The comparison of these numbers reflects the fraction
 of the background events that were subtracted from
 the reaction candidates.
\begin{table*}
\caption
[tab:events]{
 Number of the initial candidates, $N_{\mathrm{Cand}}$,
 for the radiative reactions $K^-p \to \gamma \Lambda$
 and $K^-p \to \gamma \Sigma^0$
 and the events, $N_{\mathrm{Evnt}}$, left after the background
 subtraction for the eight beam momenta. The last two rows
 list the corresponding numbers for $K^-p \to \gamma \Sigma^0$ as
 reported in Ref.~\protect\cite{SGam}.
 } \label{tab:events}
\begin{ruledtabular}
\begin{tabular}{|l|c|c|c|c|c|c|c|c|} 
\hline
 $p_{K^-}\pm \delta_p$ (MeV/$c$) &
 514$\pm$10 & 560$\pm$11 & 581$\pm$12 & 629$\pm$11 &
 659$\pm$12 & 687$\pm$11 & 714$\pm$11 & 750$\pm$13 \\
\hline
 $N_{\mathrm{Cand}}(K^- p \to \gamma \Lambda)$ & 567 & 1103 & 1691 & 1785 
                           & 1836 & 2099 & 2273 & 3456 \\
 $N_{\mathrm{Evnt}}(K^- p \to \gamma \Lambda)$ & 140 & 236 & 301 & 454 
                           & 344 & 453 & 491 & 739 \\
 $N_{\mathrm{Cand}}(K^- p \to \gamma \Sigma^0)$ & 249 & 486 & 751 & 838 
                           & 977 & 1226 & 1331 & 2407 \\
 $N_{\mathrm{Evnt}}(K^- p \to \gamma \Sigma^0)$ & 67 & 130 & 248 & 319 
                           & 339 & 430 & 581 & 1120 \\
\hline
 $N_{\mathrm{Cand}}(K^- p \to \gamma \Sigma^0)$, V-A & 35 & 87 & 139 & 216 
                           & 234 & 383 & 576 & 647 \\
 $N_{\mathrm{Evnt}}(K^- p \to \gamma \Sigma^0)$, V-A & 1 & 8 & 4 & 33 
                           & 10 & 33 & 196 & 214 \\
\hline
\end{tabular}
\end{ruledtabular}
\end{table*}

 The differential cross sections are given as a function of $\cos\theta^*$,
 where the full range from $-1$ to 1 is divided into 12 bins.
 Since the $K^-p \to \gamma \Lambda$ reaction for the four highest
 beam momenta has very low acceptance in the last bin, the
 results for this bin are not presented.
 To calculate cross sections,
 the Particle Data Group (PDG)~\cite{PDG} branching ratio $0.358\pm 0.005$
 for the $\Lambda \to \pi^0 n$ decay was used.
 The effective proton density in the target times
 the effective target length was
 $(4.05\pm0.08)\times 10^{-7}$~ $\mu$b$^{-1}$. 
 The calculation of the total number
 of beam kaons incident on the target
 is given in detail in Ref.~\cite{NakPhD}.
 The uncertainties presented in our results for the differential
 cross sections are statistical only.

 The systematic uncertainty that includes the evaluation of
 the losses of good events due to pileup clusters
 and the uncertainty in
 the total number of beam kaons incident on the target
 is about 7\%. This uncertainty is the same for all
 $K^- p\to \mathrm{neutrals}$ reactions and has been determined
 in the earlier analyses of the data~\cite{k0sn_lpi0_spi0}.
 Another systematic uncertainty comes from the subtraction
 of large backgrounds from other reactions, the shapes
 and contributions of which are based on the MC simulation of those
 reactions. This uncertainty was estimated
 by varying the kinematic-fit CL value used for event selection,
 where a tighter cut on the CL leads to a better ratio
 of the signal to the background. In addition to our
 standard 5\%-CL cut, the 10\% and
 20\% cuts were also tested. All results agreed within the statistical
 uncertainties. The largest fluctuations were observed
 in the first and the last bins, which have the smallest
 acceptance. No preferred trend of the results to smaller
 or larger values after tightening the CL cut was observed.
 Our conservative value for this type of
 systematic uncertainty is 7\%. All imperfections
 in the shapes of our differential cross sections can
 be explained by statistical fluctuations.
 Adding our two systematic uncertainties in quadrature
 gives 10\% for our total systematic uncertainty; it
 is not included in the errors presented
 in the tables and figures with our results. 

\begin{figure*}
\includegraphics[width=17.cm,height=10.cm,bbllx=1.cm,bblly=1.cm,bburx=19.5cm,bbury=11.cm]{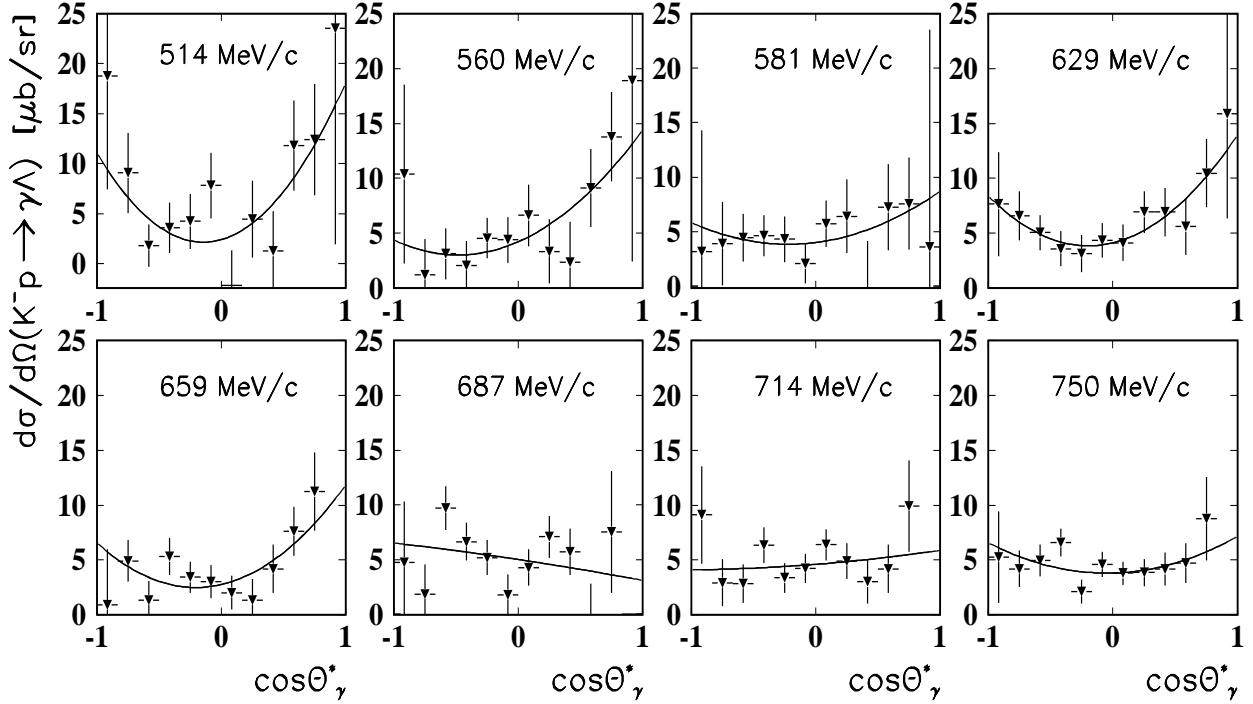}
\caption{
 Our differential cross sections for $K^-p \to \gamma \Lambda$
 at the eight beam momenta.
 The curves are the Legendre polynomial fits to our data. 
}
 \label{fig:lamgam_dxs_lpl} 
\end{figure*}

\begin{figure*}
\includegraphics[width=17.cm,height=10.cm,bbllx=1.cm,bblly=1.cm,bburx=19.5cm,bbury=11.cm]{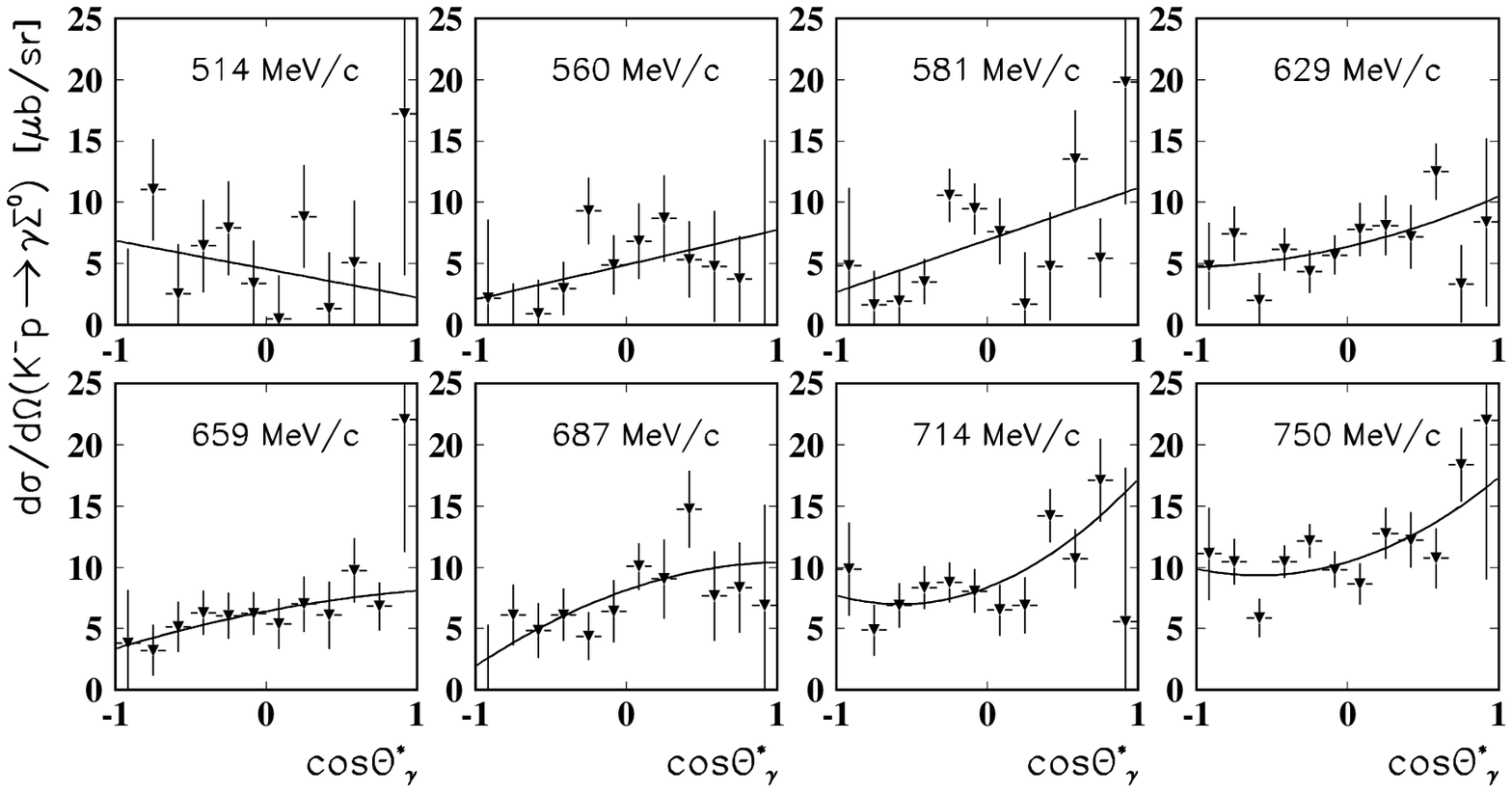}
\caption{
 Our differential cross sections for $K^-p \to \gamma \Sigma^0$
 at the eight beam momenta.
 The curves are the Legendre polynomial fits to our data. 
}
 \label{fig:siggam_dxs_lpl} 
\end{figure*}

 Our differential cross sections for $K^- p \to \gamma \Lambda$
 as a function of $\cos\theta^*$ for the outgoing photon
 are given for each of the eight beam momenta in
 Tables~\ref{tab:lgam1} and \ref{tab:lgam2}. 
 The corresponding results for $K^- p \to \gamma \Sigma^0$
 are given in Tables~\ref{tab:sgam1} and \ref{tab:sgam2}.
 Because of our limited statistics and the subtraction
 of the large backgrounds, several points resulted in
 unphysical negative values; however, all of them are consistent
 with positive values within their uncertainties.
\begin{table*}
\caption
[tab:lgam1]{
 Differential cross sections for the $K^- p \to \gamma \Lambda$ reaction
 for the four lowest beam momenta.
 } \label{tab:lgam1}
\begin{ruledtabular}
\begin{tabular}{|c|c|c|c|c|} 
\hline
 $p_{K^-}$ (MeV/$c$) & $514\pm 10$ & $560\pm 11$ &
                                     $581\pm 12$ & $629\pm 11$ \\
\hline
  $\cos\theta^{\ast}_{\gamma}$ & $d\sigma/d\Omega$ ($\mu$b/sr) &
  $d\sigma/d\Omega$ ($\mu$b/sr)&  $d\sigma/d\Omega$ ($\mu$b/sr) &  $d\sigma/d\Omega$ ($\mu$b/sr)\\
\hline
 -0.917 & $18.7\pm 11.3$ & $10.4\pm 8.2$ & $3.2\pm 11.1$ & $7.7\pm 4.8$ \\
 -0.750 & $9.1\pm 4.0$ & $1.2\pm 3.2$ & $3.9\pm 3.8$ & $6.6\pm 2.2$ \\
 -0.583 & $1.8\pm 2.1$ & $3.1\pm 2.3$ & $4.4\pm 2.2$ & $5.0\pm 1.6$ \\
 -0.417 & $3.6\pm 2.5$ & $2.1\pm 2.2$ & $4.6\pm 1.9$ & $3.6\pm 1.6$ \\
 -0.250 & $4.2\pm 2.8$ & $4.5\pm 1.9$ & $4.3\pm 2.1$ & $3.1\pm 1.7$ \\
 -0.083 & $7.8\pm 3.3$ & $4.4\pm 2.1$ & $2.0\pm 1.8$ & $4.3\pm 1.6$ \\
  0.083 & $-2.2\pm 3.5$ & $6.6\pm 2.8$ & $5.7\pm 2.2$ & $4.1\pm 1.7$ \\
  0.250 & $4.5\pm 3.8$ & $3.3\pm 2.9$ & $6.4\pm 3.4$ & $6.9\pm 1.9$ \\
  0.417 & $1.2\pm 4.0$ & $2.3\pm 3.7$ & $-0.2\pm 4.3$ & $6.9\pm 2.2$ \\
  0.583 & $11.8\pm 4.5$ & $9.1\pm 3.5$ & $7.2\pm 3.9$ & $5.6\pm 2.6$ \\
  0.750 & $12.4\pm 5.6$ & $13.8\pm 4.1$ & $7.6\pm 4.2$ & $10.5\pm 3.1$ \\
  0.917 & $23.5\pm 21.6$ & $18.9\pm 16.5$ & $3.6\pm 19.9$ & $15.9\pm 9.5$ \\
\hline
\end{tabular}
\end{ruledtabular}
\end{table*}
\begin{table*}
\caption
[tab:lgam2]{
 Differential cross sections for the $K^- p \to \gamma \Lambda$ reaction
 for the four highest beam momenta.
 } \label{tab:lgam2}
\begin{ruledtabular}
\begin{tabular}{|c|c|c|c|c|} 
\hline
 $p_{K^-}$ (MeV/$c$) & $659\pm 12$ & $687\pm 11$ &
                                     $714\pm 11$ & $750\pm 13$ \\
\hline
  $\cos\theta^{\ast}_{\gamma}$ & $d\sigma/d\Omega$ ($\mu$b/sr) &
  $d\sigma/d\Omega$ ($\mu$b/sr)&  $d\sigma/d\Omega$ ($\mu$b/sr) &  $d\sigma/d\Omega$ ($\mu$b/sr)\\
\hline
 -0.917 & $0.9\pm 5.1$ & $4.8\pm 5.5$ & $9.1\pm 4.4$ & $5.3\pm 4.2$ \\
 -0.750 & $4.9\pm 1.9$ & $1.9\pm 2.7$ & $2.9\pm 2.1$ & $4.2\pm 1.6$ \\
 -0.583 & $1.4\pm 1.7$ & $9.7\pm 2.0$ & $2.9\pm 1.8$ & $5.0\pm 1.5$ \\
 -0.417 & $5.3\pm 1.7$ & $6.6\pm 1.7$ & $6.4\pm 1.6$ & $6.6\pm 1.3$ \\
 -0.250 & $3.4\pm 1.4$ & $5.2\pm 1.6$ & $3.4\pm 1.4$ & $2.1\pm 1.1$ \\
 -0.083 & $3.0\pm 1.5$ & $1.8\pm 1.8$ & $4.3\pm 1.3$ & $4.6\pm 1.2$ \\
  0.083 & $2.0\pm 1.5$ & $4.3\pm 1.7$ & $6.4\pm 1.4$ & $3.8\pm 1.0$ \\
  0.250 & $1.4\pm 1.9$ & $7.1\pm 1.9$ & $4.9\pm 1.6$ & $3.9\pm 1.2$ \\
  0.417 & $4.2\pm 2.2$ & $5.7\pm 2.1$ & $3.0\pm 2.0$ & $4.2\pm 1.5$ \\
  0.583 & $7.6\pm 2.3$ & $0.0\pm 2.9$ & $4.2\pm 2.2$ & $4.7\pm 1.8$ \\
  0.750 & $11.2\pm 3.6$ & $7.5\pm 5.6$ & $9.9\pm 4.2$ & $8.9\pm 3.8$ \\
  0.917 & --- & --- & --- & --- \\
\hline
\end{tabular}
\end{ruledtabular}
\end{table*}
\begin{table*}
\caption
[tab:sgam1]{
 Differential cross sections for the $K^- p \to \gamma \Sigma^0$ reaction
 for the four lowest beam momenta.
 } \label{tab:sgam1}
\begin{ruledtabular}
\begin{tabular}{|c|c|c|c|c|} 
\hline
 $p_{K^-}$ (MeV/$c$) & $514\pm 10$ & $560\pm 11$ &
                                     $581\pm 12$ & $629\pm 11$ \\
\hline
  $\cos\theta^{\ast}_{\gamma}$ & $d\sigma/d\Omega$ ($\mu$b/sr) &
  $d\sigma/d\Omega$ ($\mu$b/sr)&  $d\sigma/d\Omega$ ($\mu$b/sr) &  $d\sigma/d\Omega$ ($\mu$b/sr)\\
\hline
 -0.917 & $-2.2\pm 8.4$ & $2.2\pm 6.4$ & $4.8\pm 6.3$ & $4.8\pm 3.5$ \\
 -0.750 & $11.0\pm 4.1$ & $-0.6\pm 3.9$ & $1.6\pm 2.8$ & $7.4\pm 2.2$ \\
 -0.583 & $2.6\pm 4.0$ & $0.9\pm 2.7$ & $1.9\pm 2.6$ & $2.0\pm 2.3$ \\
 -0.417 & $6.4\pm 3.8$ & $3.0\pm 2.2$ & $3.5\pm 1.9$ & $6.2\pm 1.7$ \\
 -0.250 & $7.9\pm 3.8$ & $9.3\pm 2.7$ & $10.5\pm 2.2$ & $4.3\pm 1.7$ \\
 -0.083 & $3.4\pm 3.5$ & $4.9\pm 2.4$ & $9.5\pm 2.1$ & $5.7\pm 1.6$ \\
  0.083 & $0.5\pm 3.6$ & $6.8\pm 3.1$ & $7.6\pm 2.7$ & $7.8\pm 2.2$ \\
  0.250 & $8.8\pm 4.2$ & $8.7\pm 3.5$ & $1.7\pm 4.2$ & $8.1\pm 2.4$ \\
  0.417 & $1.3\pm 4.5$ & $5.3\pm 3.1$ & $4.8\pm 4.4$ & $7.2\pm 2.6$ \\
  0.583 & $5.1\pm 5.1$ & $4.8\pm 4.5$ & $13.5\pm 4.0$ & $12.5\pm 2.3$ \\
  0.750 & $-0.4\pm 5.5$ & $3.8\pm 3.5$ & $5.5\pm 3.2$ & $3.3\pm 3.2$ \\
  0.917 & $17.2\pm 13.2$ & $-0.8\pm 15.9$ & $19.8\pm 9.9$ & $8.4\pm 6.9$ \\
\hline
\end{tabular}
\end{ruledtabular}
\end{table*}
\begin{table*}
\caption
[tab:sgam2]{
 Differential cross sections for the $K^- p \to \gamma \Sigma^0$ reaction
 for the four highest beam momenta.
 } \label{tab:sgam2}
\begin{ruledtabular}
\begin{tabular}{|c|c|c|c|c|} 
\hline
 $p_{K^-}$ (MeV/$c$) & $659\pm 12$ & $687\pm 11$ &
                                     $714\pm 11$ & $750\pm 13$ \\
\hline
  $\cos\theta^{\ast}_{\gamma}$ & $d\sigma/d\Omega$ ($\mu$b/sr) &
  $d\sigma/d\Omega$ ($\mu$b/sr)&  $d\sigma/d\Omega$ ($\mu$b/sr) &  $d\sigma/d\Omega$ ($\mu$b/sr)\\
\hline
 -0.917 & $3.8\pm 4.3$ & $-2.1\pm 7.5$ & $9.9\pm 3.8$ & $11.1\pm 3.8$ \\
 -0.750 & $3.2\pm 2.1$ & $6.1\pm 2.5$ & $4.9\pm 2.1$ & $10.4\pm 1.9$ \\
 -0.583 & $5.1\pm 2.0$ & $4.8\pm 2.3$ & $6.9\pm 1.8$ & $5.9\pm 1.6$ \\
 -0.417 & $6.3\pm 1.8$ & $6.1\pm 2.1$ & $8.3\pm 1.8$ & $10.5\pm 1.3$ \\
 -0.250 & $6.1\pm 1.9$ & $4.4\pm 2.0$ & $8.8\pm 1.6$ & $12.1\pm 1.4$ \\
 -0.083 & $6.2\pm 1.8$ & $6.4\pm 2.5$ & $8.1\pm 1.8$ & $9.8\pm 1.5$ \\
  0.083 & $5.4\pm 2.1$ & $10.1\pm 1.9$ & $6.5\pm 2.1$ & $8.6\pm 1.7$ \\
  0.250 & $7.0\pm 2.3$ & $9.1\pm 3.2$ & $6.9\pm 2.3$ & $12.8\pm 2.1$ \\
  0.417 & $6.1\pm 2.7$ & $14.8\pm 3.1$ & $14.2\pm 2.2$ & $12.2\pm 2.3$ \\
  0.583 & $9.8\pm 2.6$ & $7.7\pm 3.7$ & $10.7\pm 2.4$ & $10.7\pm 2.5$ \\
  0.750 & $6.8\pm 2.0$ & $8.3\pm 3.7$ & $17.1\pm 3.4$ & $18.4\pm 3.0$ \\
  0.917 & $22.1\pm 10.8$ & $6.9\pm 8.2$ & $5.6\pm 12.5$ & $22.0\pm 12.9$ \\
\hline
\end{tabular}
\end{ruledtabular}
\end{table*}
 In Figs.~\ref{fig:lamgam_dxs_lpl} and \ref{fig:siggam_dxs_lpl},
 we show the Legendre polynomial fits
 to our differential cross sections,
\begin{equation}
 d\sigma/d\Omega=\sum_{l=0}^{l_{\mathrm{max}}}A_lP_l(\cos\theta^*),
\label{eqn:legpol}
\end{equation}
 where $P_l$ is the Legendre polynomial of order $l$, and $A_l$ is
 its coefficient. The choice of the maximum order
 $l_{\mathrm{max}}$ was limited by our statistics:
 it was 1 for the three lowest beam momenta
 of the $K^- p \to \gamma \Sigma^0$ reaction and 2 for all other fits.
 The results of Legendre polynomial fits are given
 for both the reactions in
 Tables~\ref{tab:ftlgam} and \ref{tab:ftsgam}.    
\begin{table*}
\caption
[tab:ftsgam]{
 Legendre polynomial coefficients for 
 the $K^- p \to \gamma \Lambda$ reaction at the eight beam momenta.
 } \label{tab:ftlgam}
\begin{ruledtabular}
\begin{tabular}{|c|c|c|c|c|} 
\hline
 $p_{K^-}$ (MeV/$c$) & $A_0$ & $A_1$ & $A_2$ & $\chi^2$/ndf \\
\hline
 $514\pm 10$ & $6.45\pm 1.39$ & $3.49\pm 2.51$ & $8.11\pm 3.87$ & 1.13 \\
 $560\pm 11$ & $5.94\pm 1.12$ & $5.01\pm 2.07$ & $3.44\pm 2.91$ & 0.59 \\
 $581\pm 12$ & $5.08\pm 1.21$ & $1.44\pm 2.17$ & $2.18\pm 3.05$ & 0.43 \\
 $629\pm 11$ & $6.46\pm 0.78$ & $2.77\pm 1.46$ & $4.68\pm 2.05$ & 0.27 \\
 $659\pm 12$ & $4.90\pm 0.80$ & $2.61\pm 1.48$ & $4.29\pm 2.06$ & 1.04 \\
 $687\pm 11$ & $4.95\pm 1.00$ & $-1.70\pm 1.79$ & $-0.14\pm 2.62$ & 1.83 \\
 $714\pm 11$ & $4.74\pm 0.81$ & $0.89\pm 1.49$ & $0.25\pm 2.07$ & 1.03 \\
 $750\pm 13$ & $4.83\pm 0.67$ & $0.30\pm 1.21$ & $2.03\pm 1.72$ & 1.01 \\
\end{tabular}
\end{ruledtabular}
\end{table*}
\begin{table*}
\caption
[tab:ftsgam]{
 Legendre polynomial coefficients for 
 the $K^- p \to \gamma \Sigma^0$ reaction at the eight beam momenta.
 } \label{tab:ftsgam}
\begin{ruledtabular}
\begin{tabular}{|c|c|c|c|c|} 
\hline
 $p_{K^-}$ (MeV/$c$) & $A_0$ & $A_1$ & $A_2$ & $\chi^2$/ndf \\
\hline
 $514\pm 10$ & $4.55\pm 1.30$ & $-2.32\pm 2.81$ & --- & 0.83 \\
 $560\pm 11$ & $4.90\pm 0.95$ & $2.84\pm 2.11$ & --- & 0.75 \\
 $581\pm 12$ & $6.91\pm 0.89$ & $4.26\pm 1.93$ & --- & 1.59 \\
 $629\pm 11$ & $6.78\pm 0.78$ & $2.88\pm 1.49$ & $0.81\pm 1.96$ & 1.25 \\
 $659\pm 12$ & $6.18\pm 0.71$ & $2.36\pm 1.34$ & $-0.46\pm 1.90$ & 0.42 \\
 $687\pm 11$ & $7.49\pm 0.98$ & $4.24\pm 1.85$ & $-1.31\pm 2.53$ & 0.84 \\
 $714\pm 11$ & $9.72\pm 0.79$ & $4.73\pm 1.51$ & $2.73\pm 2.09$ & 1.07 \\
 $750\pm 13$ & $11.50\pm 0.75$ & $3.72\pm 1.40$ & $2.11\pm 1.89$ & 1.55 \\
\hline
\end{tabular}
\end{ruledtabular}
\end{table*}

\begin{figure*}
\includegraphics[width=17.cm,height=10.cm,bbllx=1.cm,bblly=1.cm,bburx=19.5cm,bbury=11.cm]{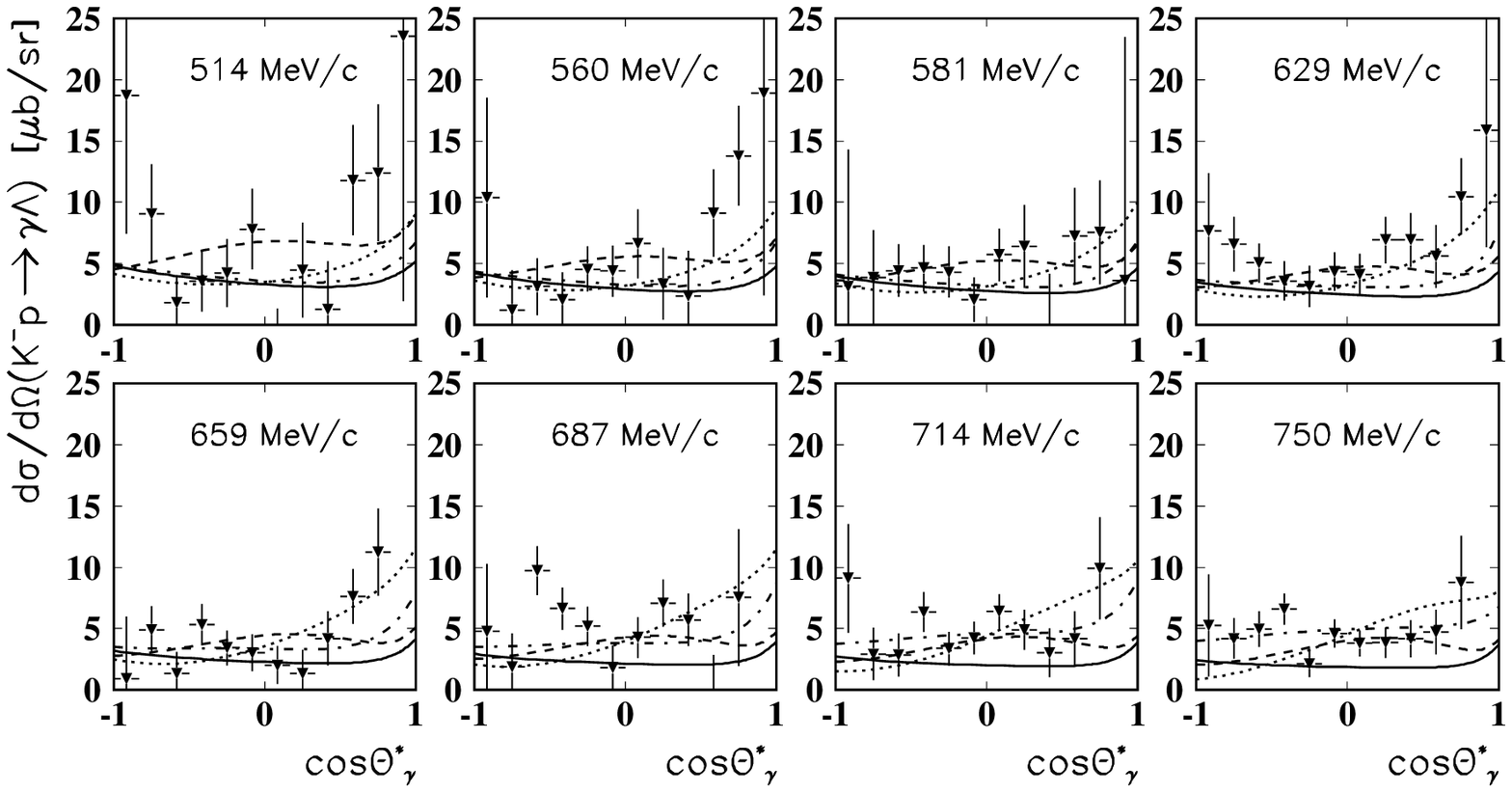}
\caption{
 Comparison of our differential cross sections
 for $K^-p \to \gamma \Lambda$ with RPR model calculations.
 The solid line is the Regge-2 model prediction consisting
 solely of non-resonant diagrams.
 The dash, dot, and dash-dot curves represent fits to our data of
 the Regge-2 model supplemented with the $\Lambda(1520)D_{03}$,
 $\Sigma(1660)P_{11}$, or $\Sigma(1670)D_{13}$ resonance, respectively.
}
 \label{fig:lamgam_dxs_gent} 
\end{figure*}

\begin{figure*}
\includegraphics[width=17.cm,height=10.cm,bbllx=1.cm,bblly=1.cm,bburx=19.5cm,bbury=11.cm]{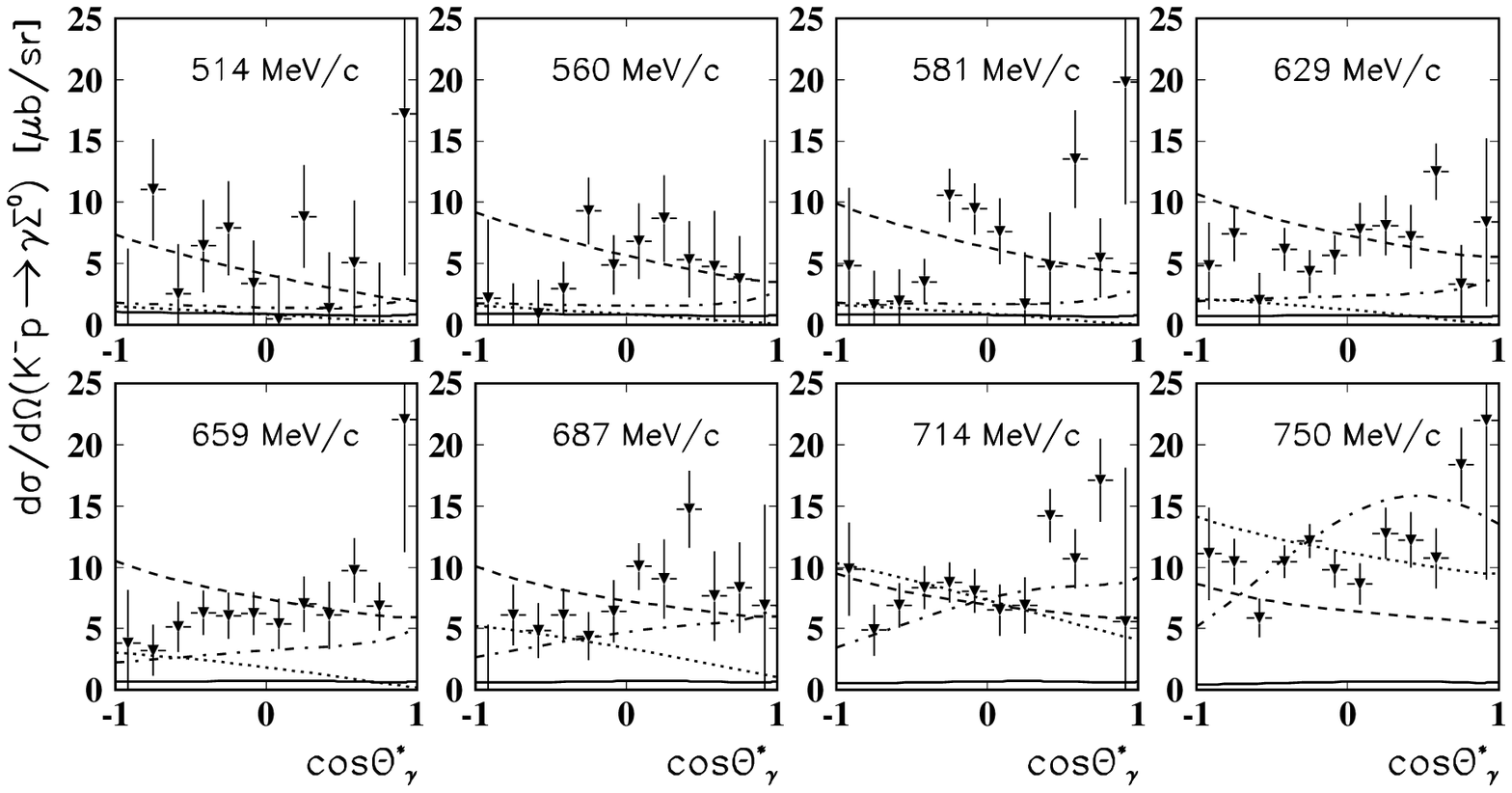}
\caption{
 Comparison of our differential cross sections
 for $K^-p \to \gamma \Sigma^0$ with RPR model calculations.
 The solid line is the Regge-3 model prediction consisting
 solely of non-resonant diagrams.
 The dash, dot, and dash-dot curves represent fits to our data of
 the Regge-3 model supplemented with the $\Lambda(1600)P_{01}$,
 $\Lambda(1670)S_{01}$, or $\Lambda(1690)D_{03}$ resonance, respectively.
}
 \label{fig:siggam_dxs_gent} 
\end{figure*}

 To interpret our results for the radiative reactions
 $K^- p \to \gamma \Lambda$ and $K^- p \to \gamma \Sigma^0$, 
 we compared them with Regge-plus-resonance (RPR)
 model calculations. 
 The RPR approach has been developed to describe photo-induced
 and electro-induced kaon production off
 protons~\cite{Corthals06,Corthals07_1,Corthals07_2,Vancraeyveld09}.
 The $p(\gamma,K)Y$ amplitude, which is dominated by non-resonant
 $t$-channel contributions, is modelled in terms of $K^+(494)$
 and $K^{\ast+}(892)$ Regge-trajectory exchange.
 For the $p(\gamma,K^+)\Lambda$ reaction, the best model is coined Regge-2.
 In $K\Sigma$ photoproduction two equivalent Regge models are available.
 A recent analysis of $n(\gamma,K^+)\Sigma^-$ data allowed
 to isolate an optimal model, coined Regge-3.
 The Regge-model amplitudes can be supplemented with
 a selection of $s$-channel resonance diagrams.
 Since the radiative kaon-capture reaction is related to
 photo-induced kaon production through crossing symmetry,
 the amplitude for the kaon-capture process can be obtained
 by analytic continuation of the kaon-production amplitude
 in which the signs of the kaon and photon momenta are reversed.
 Crossing symmetry interchanges the roles of the Mandelstam-$s$
 and -$u$ variables. The Mandelstam-$t$ variable remains the same,
 as do the contributions to the amplitude that arise from
 $t$-channel exchange. Therefore, one can apply the Regge model,
 developed for kaon photoproduction, to the description of
 radiative kaon capture, without introducing or adjusting
 any parameters~\cite{VanCauteren09}.
 As can be appreciated in Fig.~\ref{fig:lamgam_dxs_gent},
 the Regge-model predictions (solid lines) for $K^-p \to \gamma\Lambda$
 are of the same order as the measured differential cross sections
 and are in reasonable agreement with the data except for
 an underprediction of the strength at forward angles.
 The situation for the $K^-p\to \gamma\Sigma^0$ channel
 is entirely different. One can see in Fig.~\ref{fig:siggam_dxs_gent}
 that the Regge model underpredicts the data by an order of
 magnitude approximately. This hints at an important role
 for resonance-exchange terms in the reaction amplitude.

 In order to assess the possible resonant contributions
 to the radiative capture reactions, the Regge-model amplitudes
 can be enriched with hyperon-resonance exchange terms,
 along similar lines as $s$-channel resonances have been added
 for the $p(\gamma,K)Y$ processes~\cite{Corthals06,Corthals07_1}.
 Table~\ref{tab:hyperonResonances} lists six established hyperon
 resonances relevant to the energy range of the data presented here.
 Due to the lack of experimental data for the electromagnetic
 decay widths of these resonances, we have included predictions
 by the Bonn constituent-quark
 model~\cite{VanCauteren05,VanCauteren07}.
 When fixing the resonances' mass and width at the central RPP value,
 inclusion of a spin-1/2 or spin-3/2 resonance
 introduces two or three free parameters
 respectively\pone\footnotetext[1]{The most general Lagrangian for spin-3/2
 resonances has three additional degrees-of-freedom,
 often called \textit{off-shell} parameters.
 To limit the total number of fitting parameters,
 we discard the latter.}.
 The quality and amount of the differential-cross-section data
 presented here are not sufficient to perform fits with
 multiple resonances. Therefore, we carried out fits with
 a single resonance at a time. For the $K^-p \to \gamma\Lambda$
 channel, we obtain a typical $\chi^2$ value of 115
 for 92 data points. Five of the six fits
 for the $K^-p \to \gamma\Sigma^0$ reaction
 attain $\chi^2$ of 230 for 96 data points,
 except for the fit including $\Lambda(1670)S_{01}$,
 which converged at a $\chi^2$ value of 355 for 96 data points.

 In Fig.~\ref{fig:lamgam_dxs_gent}, we show the $K^-p \to \gamma\Lambda$
 model calculations including the $\Lambda(1520)D_{03}$,
 $\Sigma(1660)P_{11}$, or $\Sigma(1670)D_{13}$ state,
 since, according to the Bonn model, these resonances have
 the largest branching ratio to the $\gamma\Lambda$ channel.
 The strength of the $\Sigma(1660)P_{11}$ and $\Sigma(1670)D_{13}$
 resonances increases with the energy of the incoming kaon,
 whereas the $\Lambda(1520)D_{03}$ contribution is more
 uniformly spread. The angular dependence of these three models
 are in reasonable agreement with our data.
 Figure~\ref{fig:siggam_dxs_gent} features model
 calculations for the $K^-p \to \gamma\Sigma^0$ channel
 including the $\Lambda(1600)P_{01}$, $\Lambda(1670)S_{01}$, or
 $\Lambda(1690)D_{03}$ resonance.
 The angular dependences of $\Lambda(1600)P_{01}$ and
 $\Lambda(1670)S_{01}$ are similar, but the strength of
 the $\Lambda(1600)P_{01}$ contribution is much more uniform
 as a function of the beam energy.
 The angular dependence of $\Lambda(1690)D_{03}$ has a
 trend opposite to $\Lambda(1600)P_{01}$ and $\Lambda(1670)S_{01}$ and
 better matches the experimental data at the highest momenta. 
 
\begin{figure}
\includegraphics[width=7.cm,height=10.cm,bbllx=1.cm,bblly=0.5cm,bburx=14.cm,bbury=19.cm]{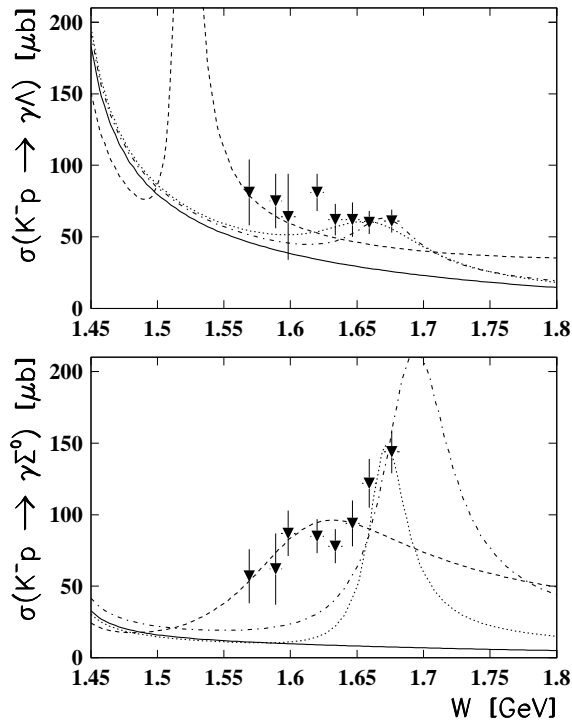}
\caption{
 Our total cross sections as a function of
 the center-of-mass energy $W$
 for $K^-p \to \gamma \Lambda$ (top)
 and $K^-p \to \gamma \Sigma^0$ (bottom).
 The solid lines are the Regge-2 model prediction for
 $K^-p \to \gamma \Lambda$ (top) and the Regge-3 model prediction
 for $K^-p \to \gamma \Sigma^0$ (bottom).
 The dash, dot, and dash-dot curves shown for $K^-p \to \gamma \Lambda$
 represent the Regge-2 model supplemented with the $\Lambda(1520)D_{03}$,
 $\Sigma(1660)P_{11}$, or $\Sigma(1670)D_{13}$ resonance, respectively.
 The dash, dot, and dash-dot curves shown for $K^-p \to \gamma \Sigma^0$
 represent the Regge-3 model supplemented with
 the $\Lambda(1600)P_{01}$, $\Lambda(1670)S_{01}$, or
 $\Lambda(1690)D_{03}$ resonance, respectively.
}
 \label{fig:lgam_sgam_xst} 
\end{figure}

 The evaluation of the total cross sections for
 the radiative reactions $K^-p \to \gamma \Lambda$
 and $K^- p \to \gamma\Sigma^0$  
 was based on the Legendre polynomial fits
 shown in Figs.~\ref{fig:lamgam_dxs_lpl} and
 \ref{fig:siggam_dxs_lpl}. The results obtained
 for both radiative reactions are listed
 for each of the eight beam momenta in
 Table~\ref{tab:totxsc} and are also shown
 in Fig.~\ref{fig:lgam_sgam_xst} in conjunction
 with the RPR model calculations.
 We assume that the systematic uncertainties in our
 values for the total cross sections have the same magnitude
 as for the differential cross sections.
 These systematic uncertainties are not included
 into the presented results. 
\begin{table*}
\caption
[tab:totxsc]{
 The total cross sections for the radiative reactions
 $K^- p \to \gamma \Lambda$ and 
 $K^- p \to \gamma \Sigma^0$ at the eight beam momenta.
 } \label{tab:totxsc}
\begin{ruledtabular}
\begin{tabular}{|l|c|c|c|c|c|c|c|c|} 
\hline
 $p_{K^-}$ (MeV/$c$)&
 $514\pm 10$ & $560\pm 11$ & $581\pm 12$ & $629\pm 11$ &
 $659\pm 12$ & $687\pm 11$ & $714\pm 11$ & $750\pm 13$ \\
 $W$ (MeV)&
 $1569\pm 4$ & $1589\pm 5$ & $1598\pm 5$ & $1620\pm 5$ &
 $1634\pm 5$ & $1647\pm 5$ & $1659\pm 5$ & $1676\pm 6$  \\
\hline
 $\sigma_{\gamma \Lambda}$ ($\mu$b) & $81\pm 23$ & $75\pm 19$ & $64\pm 30$
  & $81\pm 13$ & $62\pm 11$ & $62\pm 12$ & $60\pm 8$ & $61\pm 8$ \\
 $\sigma_{\gamma \Sigma^0}$ ($\mu$b) & $57\pm 19$ & $62\pm 25$ & $87\pm 16$
  & $85\pm 12$ & $78\pm 12$ & $94\pm 16$ & $122\pm 17$ & $144\pm 15$ \\
\hline
\end{tabular}
\end{ruledtabular}
\end{table*}

 For the $K^-p\rightarrow \gamma\Lambda$ reaction,
 the total cross section falls off as the
 center-of-mass energy $W$ rises.
 This trend is predicted by the Regge model, which accounts
 for roughly half the strength.
 The addition of the $\Lambda(1520)D_{03}$ resonance allows
 to largely make up for the missing strength.
 The Bonn constituent-quark model also predicts a large
 electromagnetic decay width for the transitions of
 $\Sigma(1660)P_{11}$ and $\Sigma(1670)D_{13}$ to
 $\Lambda(1116)\gamma$. The RPR fit including any of
 these two resonances
 improves the description of the total cross section
 at the highest energy bins, but fails to account for
 the rise at lower energies.
 In the $K^-p \to \gamma\Sigma^0$ reaction channel,
 the energy dependence of the total cross section differs
 notably from the $\gamma\Lambda$ final state,
 peaking in the highest measured energy bin.
 This behavior is opposite to the Regge-model result,
 which underestimates the total cross section by a factor
 of four and predicts a fall off as energy increases.
 A nice correspondence with the data at lower energies
 is realized through the inclusion of
 the $\Lambda(1600)P_{01}$ resonance, which has
 a particularly large value for the total decay width in the RPP.
 The Regge-model calculations supplemented with either
 the $\Lambda(1670)S_{01}$ resonance or $\Lambda(1670)S_{01}$,
 having a large value for $\Gamma_{\gamma\Sigma^0}$ in the Bonn model,
 allow to reproduce the apparent peak in the total cross section data,
 but fail at lower energies. For both reaction channels
 presented in this work, the data cannot be understood in terms
 of a reaction amplitude consisting of non-resonant terms
 in conjunction with a single resonance. The energy dependence
 of the total cross sections, however, suggests
 a prominent contribution from a resonance in the range
 of 1550~MeV for the $\gamma\Lambda$ final state and
 of 1700~MeV for $\gamma\Sigma^0$. 
 
\section{Comparison with other results}

\begin{figure}
\includegraphics[width=7.cm,height=7.cm,bbllx=1.cm,bblly=0.5cm,bburx=13.cm,bbury=13.cm]{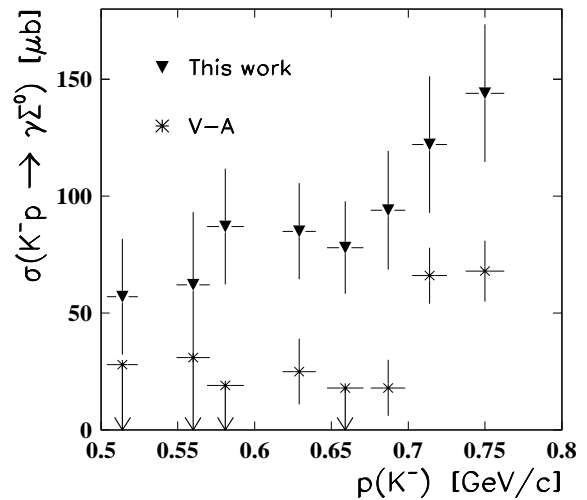}
\caption{
 Our $K^-p \to \gamma \Sigma^0$ total cross sections
 as a function of the beam momentum compared to
 the V-A results from Ref.~\protect\cite{SGam}.
 The uncertainties plotted for our results are the 
 sum of statistical and systematic uncertainties.
 The uncertainties plotted for V-A results are statistical
 only. The V-A points shown by arrows are upper limits.
}
 \label{fig:sgam_va_xst} 
\end{figure}

 Our results for $K^- p \to \gamma\Lambda$ in flight are
 unique and at the moment can be compared
 to theoretical predictions only.
 For the $K^- p \to \gamma\Sigma^0$ reaction,
 there is also an independent analysis of the
 same data set, resulting only in the determination
 of the total cross sections~\cite{SGam}.
 That analysis was conducted by the Valparaiso-Argonne (V-A)
 group and published before the present analysis became
 available.
 We now believe that the V-A analysis has been published
 prematurely because the comparison of the two analyses
 (see Fig.~\ref{fig:sgam_va_xst}) reveals that the V-A
 results for the total cross sections are systematically
 lower than ours. The magnitude of the disagreement
 between the two analyses cannot be evaluated accurately
 because of the inadequate treatment of the systematic
 uncertainties in the V-A analysis.
 Moreover, for half of the V-A cross sections, only upper limits
 were reported. Those upper limits are based on the statistical
 uncertainties only and the confidence level at which they
 have been determined is omitted in Ref.~\cite{SGam}.
 Our total cross sections are plotted in Fig.~\ref{fig:sgam_va_xst}
 after adding linearly our statistical and systematic uncertainties.

 The difference between our total cross sections
 and the V-A ones is of the same order of magnitude
 ($50-75~\mu$b) for all beam momenta.
 Such a feature is indicative of the oversubtraction
 of the large background contributions
 in the V-A analysis. At our energies, the yield of
 the major background reactions
 ($K^-p\to \pi^0\Sigma^0$ and $K^-p\to \pi^0\Lambda$)
 is almost the same as a function of
 beam momentum~\cite{k0sn_lpi0_spi0}.
 Then a wrong normalization
 of these background contributions should result in
 such a deviation from the actual cross section,
 the magnitude of which is similar at each momentum.
 According to Ref.~\cite{SGam}, the normalization
 of the MC simulations for the background reactions
 was based on their total cross sections.
 However, it is not clear how such a normalization is
 possible if the total cross sections
 for $K^- p \to \gamma\Sigma^0$ have not yet been determined.
 In our opinion, the approach used in Ref.~\cite{SGam} to normalize
 the contributions from the background reactions led
 to their substantial oversubtraction.

 Another feature of the V-A analysis that should result in
 large systematic uncertainties
 is the use of very tight selection
 cuts to improve the signal-to-background ratio.
 To illustrate the consequences of applying such tight cuts,
 we list the numbers of the events selected
 in each of the two analyses in Table~\ref{tab:events}
 for comparison.
 The corresponding numbers of the events selected
 in the V-A analysis are many times smaller than ours. 
 Besides leading to poor statistics, the use of tight selection
 cuts requires excellent agreement between
 the experimental data and the MC simulations
 for both the signal and background reactions. 
 Besides knowing the reaction dynamics, this agreement assumes
 that the MC simulation reproduces correctly the experimental
 resolutions and the trigger conditions.
 In our opinion, such an agreement was not sufficiently
 illustrated in Ref.~\cite{SGam}, especially for
 the $\chi^2$-type function $F$ used for the selection
 of $K^- p \to \gamma\Sigma^0$ candidates.
 As all variances $1/w$, which reflect the experimental
 resolution of each term in the function $F$, have been determined
 from the MC simulation, it does not mean that they are
 correct for the experimental data.

 The sensitivity of the results to the cuts being applied
 is tested in the V-A analysis by adding those cuts one by one.
 Most of those cuts just duplicate one of the terms
 of their function $F$. It seems that instead it would be
 more reasonable to tighten only the cut on the magnitude of $F$
 for improving the signal-to-background ratio and testing
 the sensitivity of the results to that.

 The systematic uncertainties reported in Ref.~\cite{SGam}
 are, in our opinion, significantly underestimated,
 since neither the effect from very tight cuts nor
 the effect from the subtraction of large background
 contributions were taken into account.
 The overall 10\% systematics uncertainty borrowed
 from the V-A analysis of the $K^-p\to \pi^0\Sigma^0$
 reaction~\cite{SPi0} definitely does not take
 those effects into account.
 The systematic uncertainty in the acceptance
 due to a possible deviation of the $K^- p \to \gamma\Sigma^0$
 differential cross sections from the isotropic distribution
 was estimated in Ref.~\cite{SGam} as $+20\%$.
 This shows the strong
 sensitivity of the V-A results to the reaction dynamics.
 However, the approach used for that in Ref.~\cite{SGam}
 needs a better justification. The statistics available
 in the V-A analysis at the beam momentum of 750~MeV/$c$
 was sufficient for the determination of their own
 $K^- p \to \gamma\Sigma^0$ differential cross section.
 Then the difference between their standard method and
 the result obtained by the integration of the
 differential cross section would allow a better estimate
 of this type of systematic uncertainty.   
  
\section{Summary}

 For the first time, the differential cross sections
 for $K^-$ radiative capture in flight on the proton,
 leading to the $\gamma\Lambda$ and $\gamma\Sigma^0$
 final states, have been measured
 at eight $K^-$ momenta between 514 and 750~MeV/$c$.
 The data were obtained with
 the Crystal Ball multiphoton spectrometer installed
 at the separated $K/\pi$ beam line C6
 of the BNL Alternating Gradient Synchrotron.
 The results substantially improve
 the existing experimental data available
 for studying radiative decays of excited hyperon states.
 An exploratory theoretical analysis is performed
 within the Regge-plus-resonance approach keeping fixed
 the non-resonant contributions to the reaction amplitude,
 optimized against kaon photoproduction data.
 According to this analysis, the $K^- p \to \gamma\Sigma^0$
 reaction is dominated by hyperon-resonance exchange and hints
 at an important role for a resonance
 in the mass region of 1700~MeV.
 For the $K^- p \to \gamma\Lambda$ reaction, on the other hand,
 the resonant contributions account for only half the strength,
 and the data suggest the importance of a resonance
 in the mass region of 1550~MeV.

 In our opinion, the analysis of the $K^- p \to \gamma\Sigma^0$
 reaction reported in Ref.~\cite{SGam} was conducted
 with serious shortcomings, and we
 do not recommend those results to be used.
 The discrepancy between the V-A results and ours cannot
 be evaluated accurately for lack of the adequate
 systematic uncertainties in the V-A analysis.

\begin{acknowledgments}
 This work was supported in part by DOE and NSF of the U.S.,
 the Research Foundation - Flanders (FWO) of Belgium,
 the research council of Ghent University,
 NSERC of Canada, the Russian Ministry of Industry, 
 Science and Technologies, and the Russian Foundation for Basic Research.
 We thank SLAC for the loan of the Crystal Ball.
 The assistance of BNL and AGS with the setup is much appreciated.
\end{acknowledgments}

\end{document}